\documentclass[11pt]{article}

\usepackage{natbib}
\usepackage{adjustbox}
\usepackage{graphicx}
\usepackage{multirow}
\usepackage{float}
\usepackage{amssymb}

\usepackage[a4paper=true,pagebackref=true]{hyperref}
\hypersetup{colorlinks = true, linkcolor = green, anchorcolor = red, citecolor = blue, filecolor = red, pagecolor = red, urlcolor = red}

\usepackage{booktabs}
\usepackage{amsmath}
\usepackage{array}
\usepackage{caption}
\usepackage{ctable}
\DeclareMathOperator{\sech}{sech}

\title{\large\bf General Relativistic Calculations for White Dwarf Stars}

\author{Arun Mathew\thanks{a.mathew@iitg.ernet.in} \ and Malay K. Nandy\thanks{mknandy@iitg.ernet.in}\\
\normalsize Department of Physics, Indian Institute of Technology Guwahati,\\
\normalsize Guwahati 781 039, India.}

\date{}

\begin{document} 

\maketitle

\abstract{The mass-radius relations for white dwarf stars are investigated by solving the Newtonian as well as Tolman-Oppenheimer-Volkoff (TOV) equations for hydrostatic equilibrium assuming the electron gas to be non-interacting. We find that the Newtonian limiting mass of $1.4562M_\odot$  is modified to $1.4166M_\odot$ in the general relativistic case for $^4_2$He (and $^{12}_{\ 6}$C) white dwarf stars. Using the same general relativistic treatment, the critical mass for $^{56}_{26}$Fe white dwarf is obtained as $1.2230M_\odot$. In addition, departure from the ideal degenerate equation of state (EoS) is accounted for by considering Salpeter's EoS along with the TOV equations yielding slightly lower values for the critical masses, namely $1.4081M_{\odot}$ for $^4_2$He, $1.3916M_{\odot}$ for $^{12}_{\ 6}$C and $1.1565M_{\odot}$ for $^{56}_{26}$Fe white dwarfs. We also compare the critical densities for gravitational instability with the neutronization threshold densities to find that $^4_2$He and $^{12}_{\ 6}$C white dwarf stars are stable against neutronization with the critical values of $1.4081M_\odot$ and $1.3916M_{\odot}$, respectively. However the critical masses for $^{16}_{\ 8}$O, $^{20}_{10}$Ne, $^{24}_{12}$Mg,  $^{28}_{14}$Si, $^{32}_{16}$S and $^{56}_{26}$Fe white dwarf stars are lower due to neutronization. Corresponding to their central densities for neutronization thresholds, we obtain their maximum stable masses due to neutronization by solving the TOV equation coupled with the Salpeter EoS.
}

\section{Introduction}

Following the formulation of the Fermi-Dirac statistics, \cite{fowler1926} treated the electron gas in Sirius-B as a degenerate non-relativistic gas and found no limiting mass for the star. However, \cite{anderson1929}  and  \cite{stoner1929} considered the electron gas as relativistic and found the existence of a limiting density although their treatments were heuristic. \cite{chandra1931, chandra1931a, chandra1935, chandra1939} obtained the limiting mass as $0.91M_\odot$ initially by treating the degenerate electron gas as relativistic, and subsequently he succeeded in formulating the theory of white dwarfs to full generality. He employed Newtonian gravity and an equation of state valid for the entire range of electron velocities (including relativistic velocities) of the degenerate Fermi gas to obtain the equation of hydrostatic equilibrium. He thus obtained the equations in the form of Lane-Emden equation with index $3$ and solved the differential equations numerically to obtain the limiting mass of $1.44 M_\odot$. \cite{chandra1964} also considered the problem in the general relativistic framework to study the instability of a radially pulsating white dwarf star and obtained the critical mass as $1.4176M_\odot$. \cite{anand1965} studied the effect of rotation on a white dwarf star and showed that  the value of limiting mass increases to $1.704\,M_\odot$. Qualitative arguments given by \cite{landau1959} suggest that the inter-particle Coulomb interaction is negligible in a white dwarf star. Using the method of \cite{bohm1951}, \cite{singh1957} showed that the correction to the electron density due to electron-electron interaction is small and can be treated as negligible. On the other hand, \cite{salpeter} reconsidered the problem to account for Coulomb effects, Thomas-Fermi correction, exchange energy and correlation energy and showed that the equation of state departs measurably from the ideal degenerate case. 

In deriving the general relativistic equation of equilibrium for compact stars, \cite{tolman1939} and \cite{oppen1939} showed how the Newtonian equation of hydrostatic equilibrium is modified into what is known as Tolman-Oppenheimer-Volkoff (TOV) equation. They considered the energy-momentum tensor for a perfect fluid in the Einstein's field equation and solved for the metric for the interior of the star. This resulted in a set of three differential equations in four unknown functions, which are incomplete unless provided with the equation of state (EoS). While Tolman obtained the interior solution for a few different analytically tractable cases, Oppenheimer and Volkoff numerically solved those equations for massive neutron
cores by taking the full equation of state treating it as a non-interacting Fermi (neutron) gas.

In this paper, we consider the hydrostatic equilibrium of white dwarf stars and obtain the mass-radius relationship by solving the TOV equation. It is found, for large values of central densities, that the Newtonian limit  of $1.4562\,M_\odot$ is decreased to $1.4166\,M_\odot$ for $^4_2$He (and $^{12}_{\ 6}$C) white dwarf stars in the general relativistic treatment assuming the electron gas to be ideally degenerate. Furthermore, the critical mass for $^{56}_{26}$Fe white dwarf stars is found to be $1.2230M_\odot$ in the same formulation.
We also consider the effect of Coulomb interaction and other types of interactions by considering these Salpeter EoS in the same general relativistic formulation to obtain the critical masses of $1.4081M_{\odot}$ for $^4_2$He, $1.3916M_{\odot}$ for $^{12}_{\ 6}$C and $1.1565M_{\odot}$ for $^{56}_{26}$Fe white dwarfs.

We have also obtained the critical densities for gravitational instability directly from the solution of TOV equation coupled with Salpeter EoS and compare them with the neutronization thresholds. We find that $^4_2$He and $^{12}_{\ 6}$C white dwarf stars are stable against neutronization with the critical values of $1.4081M_\odot$ and  $1.3916M_\odot$ respectively, whereas for $^{16}_{\ 8}$O, $^{20}_{10}$Ne, $^{24}_{12}$Mg,  $^{28}_{14}$Si, $^{32}_{16}$S and $^{56}_{26}$Fe white dwarf stars, the critical masses for stability are smaller due to neutronization. For these white dwarf stars, we have also obtained the maximum stable masses due to neutronization by solving the TOV equation coupled with Salpeter EoS corresponding to their central densities for neutronization thresholds.

The rest of  the paper is organized as follows. In Section 2, we outline the derivation of TOV equation for a perfect fluid in equilibrium in general relativity. The problem of white dwarfs is taken up by considering the equation of state of cold degenerate electron gas. We also consider the case of Salpeter EoS to account for the non-ideal nature of the electron gas.  The non-linear coupled differential equations so obtained are solved numerically in Section 3, where the equations following from Newtonian gravity are also solved for the purpose of comparison. The mass-radius relationships obtained in the two cases are also compared. In Section 4, the instabilities due to gravitation and inverse beta decay are considered. The critical masses  for neutronization thresholds are computed for a few relevant stars by solving the TOV equation coupled with Salpeter EoS. The numerical results are presented with a few relevant plots and tables.

\section{General Relativistic Hydrostatic equilibrium}

The interior of a spherically symmetric star is described by $ds^2=e^\nu dt^2-e^\lambda dr^2-r^2 d\theta^2-r^2 \sin^2 \theta \, d\phi^2$,
where $\nu$  and $\lambda$ are functions of the radial distance $r$ for the static case. The matter inside the star is considered to be a perfect fluid with
energy-momentum tensor $T_{\alpha\beta}=(\varepsilon+p)u_\alpha u_\beta-pg_{\alpha\beta}$, where $p$ is the pressure and $\varepsilon=\rho c^2$ is the mass-energy density. \cite{tolman1939} and \cite{oppen1939} considered the corresponding Einstein's field equations in the interior part of the star and solved them with the boundary condition of the Schwarzschild solution in the exterior region. They obtained $e^{\nu(r)}=\left(1-\frac{G}{c^2}\frac{2M}{R}\right)\exp{\left[ -2\int_0^{p(r)} \frac{dp}{p+\varepsilon(p)}\right]}$, \  $e^{-\lambda}=1-\frac{G}{c^2}\frac{2m(r)}{r}$ \ and \ $\frac{dp}{dr}=-\frac{\varepsilon(r)+p(r)}{2}\frac{d\nu}{dr}$. These solutions lead to the well-known Tolman-Oppenheimer-Volkoff (TOV) equation, namely
\begin{equation}
\frac{dp(r)}{dr}=-\frac{G}{c^2}\frac{\varepsilon(r)+p(r)}{r(r-\frac{2G}{c^2}m(r))}  \left[m(r)+ \frac{4\pi}{c^2} p(r) r^3\right]
\label{eq:five}
\end{equation}
with
\begin{equation}
\frac{dm(r)}{dr}=  \frac{4\pi}{c^2} \varepsilon(r) r^2 
\label{eq:six}
\end{equation}

Eqs.\ (\ref{eq:five}) and (\ref{eq:six}),  together with the equation of state of matter $\varepsilon=\varepsilon(p)$, determine the hydrostatic equilibrium for an isotropic general relativistic non-rotating fluid sphere.

In a white dwarf star, the electrons can be treated as an ideal degenerate Fermi gas to a good approximation. The corresponding equation of state is given by the parametric forms
\begin{equation}
\begin{array}{c}
\varepsilon(\xi)=\frac{32 \mu_e H}{3 m_e} K \sinh^3 \frac{\xi}{4}f(\xi),\\[1em]  p(\xi)=\frac{1}{3}K\left(\sinh \xi-8 \sinh \frac{\xi}{2}+ 3\xi\right) \\[1em]
f(\xi)=1+\frac{3 m_e (\sinh\xi-\xi)}{32\mu_e H\sinh^3\frac{\xi}{4}}-\frac{m_e}{\mu_e H}
\end{array}
\label{eq:eight}
\end{equation}
where $\xi=4\sinh^{-1}(\frac{p_F}{m_ec})$, $K = \frac{\pi m_e^4 c^5}{4h^3}$,  $\mu_e=A/Z$ is the number of nucleons per electron and $H$ is the atomic mass unit. These equations are valid for all values of electron velocities, including the  extreme relativistic velocities. Substituting Eqs.\ (\ref{eq:eight}) in Eqs.\ (\ref{eq:five}) and
(\ref{eq:six}), the following differential equations are obtained.

\begin{equation}
\begin{array}{c}
\frac{dm}{dr}=4\pi \frac{32 \mu_e H}{3 m_e} \frac{K}{c^2}  r^2 \sinh^3\frac{\xi}{4}f(\xi),\\ [1em]
\frac{d\xi}{dr}=-\frac{32 \ G \mu_e H}{m_e c^2(\cosh\xi-4\cosh\frac{\xi}{2}+3) r}\left[\sinh^3\frac{\xi}{4}f(\xi)+\frac{m_e}{32 \mu_e H}(\sinh\xi-8\sinh\frac{\xi}{2}+3\xi)\right]\\
\times\left[m(r)+\frac{4\pi K(\sinh\xi-8\sinh\frac{\xi}{2}+3\xi) r^3}{ 3 c^2}\right]\left[r-\frac{2Gm(r)}{c^2}\right]^{-1}.\\
\label{eq:ten}
\end{array}
\end{equation}

The two differential equations\ (\ref{eq:ten}) are valid when the electron gas is treated as an ideal degenerate Fermi gas. A more realistic treatment must include the interactions among the electrons and the nuclei. \cite{salpeter} considered this situation and included the Coulomb effects, Thomas-Fermi correction, exchange energy and correlation energy to arrive at an EoS, given by 
\begin{equation}
\begin{array}{c}
\varepsilon_{\rm Coul}=-\frac{9}{10}\left(\frac{4}{9\pi}\right)^{1/3}\ \alpha \   Z^{2/3}\sinh^4\frac{\xi}{4}, \\
 \varepsilon_{\rm TF}=-\frac{162}{175}\left(\frac{4}{9\pi}\right)^{2/3}\ \alpha^2 \   Z^{4/3} \sinh^3\frac{\xi}{4} \cosh\frac{\xi}{4},\\
\varepsilon_{\rm ex}=-\frac{3}{128\pi} \ \alpha \ \left(12\xi\sinh\frac{\xi}{2}-16\cosh\frac{\xi}{2}-2e^{\frac{\xi}{4}}\cosh^3\frac{\xi}{4}-3\xi^2+18\right),\\
\varepsilon_{\rm corr}=\alpha^2 \   \sinh^3\frac{\xi}{4} \left(0.0115+0.031 \log_e[\alpha \ \sinh^{-1}\frac{\xi}{4}]  \right) 
\end{array}
\label{eq:esalp}
\end{equation}
in the units of $\frac{32}{3}K$, and 
\begin{equation}
\begin{array}{c}
p_{\rm Coul}=-\frac{16}{5}\left(\frac{4}{9\pi}\right)^{1/3}\ \alpha \  Z^{2/3} \sinh^4\frac{\xi}{4},\\
p_{\rm TF}=-\frac{576}{175}\left(\frac{4}{9\pi}\right)^{2/3}\ \alpha^2 \  Z^{4/3} \sinh^5\frac{\xi}{4} \sech\frac{\xi}{4},\\
p_{\rm ex}=\frac{\alpha}{2\pi} \left[\cosh\xi+8\cosh\frac{\xi}{2}-6 \xi\sinh\frac{\xi}{2}+\frac{3}{2}\xi^2-9  
 -\frac{4}{3}\tanh\frac{\xi}{4} \left(\sinh\xi-2\sinh\frac{\xi}{2}-3\xi\cosh\frac{\xi}{2}+3\xi  \right) \right],\\
p_{\rm corr}=-\frac{32}{9}0.0311 \ \alpha^2 \ \sinh^3\frac{\xi}{4}\\
\end{array}
\label{eq:psalp}
\end{equation}
in the units of $K$, where $Z$ is the number of protons and $A$ is the total number of nucleons in a nucleus.  

\section{Numerical solutions}

We shall solve the TOV equations for the two cases, namely, the ideal degenerate case and the non-ideal case, in this section.
\subsection{Ideal degenerate case}
When the electron gas is assumed to form an ideal degenerate Fermi gas, the TOV equations given by Eqs. (\ref{eq:five}) and (\ref{eq:six}) are coupled with the EoS given by Eqs. (\ref{eq:eight}). The resulting equations\ (\ref{eq:ten}) can be made dimensionless by introducing dimensionless variables $x=r/R_{\rm scale}$  and $u=m/M_{\rm scale}$. They reduce to the forms
\begin{equation}
\frac{du}{dx}=x^2\sinh^3\left(\frac{\xi}{4}\right)f(\xi)
\label{eq:twelve}
\end{equation}
\[
\frac{d\xi}{dx}=-\frac{1}{(\cosh\xi-4\cosh
\frac{\xi}{2}+3) x}
\left[\sinh^3\frac{\xi}{4}f(\xi)+\frac{m_e}{32\mu_e H}(\sinh\xi-8\sinh\frac{\xi}{2}+3\xi)\right]
\]
\begin{equation}
\times\left[u(x)+\frac{m_e}{32\mu_e H}(\sinh\xi-8\sinh\frac{\xi}{2}+3\xi) x^3
\right]
\left[x-\frac{m_e}{16\mu_e H} \ u(x)\right]^{-1}
\label{eq:thirteen}
\end{equation}

We have chosen the values  $R_{\rm scale}= 2.3788\mu_e^{-1}\times10^8$ cm and $M_{\rm scale}=1.6475\mu_e^{-2} \times10^{32}$ g so that the pre-factors in Eqs. (\ref{eq:twelve}) and  (\ref{eq:thirteen}) are normalized to unity. As analytical solution is not possible, we employ the fourth-order Runge-Kutta scheme [\cite{press1988}] to integrate them simultaneously. Solutions to these equations are computed for several initial values of $\xi_0$ at the centre of the star. Integration is carried out from the value $u=0$, $\xi=\xi_0$  at  $x=0$ (centre) to $x=x_b$ (surface) where $\xi_b=0$ (which makes $p=0$), and $u=u_b$. The first four entries in Table~1 display the results obtained in the range $1.0\leq\xi_0\leq19.0$.

In the limit $\xi\rightarrow\infty$, Eqs.\ (\ref{eq:twelve}) and (\ref{eq:thirteen})
reduce to the simple forms
\begin{equation}
\frac{du}{dx}=\frac{3m_e}{64\mu_e H}x^2e^\xi,
\label{eq:fourteen}
\end{equation}
\begin{equation}
\frac{d\xi}{dx}=-\frac{m_e}{8\mu_e H}\frac{1}{x(x- \frac{m_e}{32 \mu_e H}\ 2u)}
\left(u+\frac{m_e}{64\mu_e H}x^3e^\xi\right).
\label{eq:fifteen}
\end{equation}
From Eqs.\ (\ref{eq:eight}), the ratio of central pressure to the central mass-energy density for the limiting case $\xi\rightarrow\infty$ turns out to be \ 
$\frac{p}{\varepsilon}=\frac{1}{3}$, which represents a sphere of fluid with infinite density and pressure at the centre.  Eqs.\ (\ref{eq:fourteen}) and (\ref{eq:fifteen}) can be solved exactly to yield $u=\frac{48}{7}\left(\frac{\mu_e H}{m_e}\right)x$  and  $ e^\xi=\frac{1024}{7} \left(\frac{\mu_e H}{m_e}\right)^2 \ \frac{1}{x^2}$. From this solution, we get the initial condition (the $\xi_0$ value) to integrate Eqs.\ (\ref{eq:twelve}) and (\ref{eq:thirteen}) from $\xi=\xi_0$ (centre) to $\xi=0$ (surface). An analysis of Eqs.\ (\ref{eq:twelve}) and (\ref{eq:thirteen}) shows that they are well approximated by Eqs.\ (\ref{eq:fourteen}) and (\ref{eq:fifteen}) for $\xi \geq 55$ for $^4_2$He white dwarfs. The EoS  $\frac{p}{\varepsilon}=\frac{1}{3}$ is also approached closely for $\xi\geq55$. The last row in Table~1 corresponds to $\xi_0=55$.

\begin{table}[h]
\caption{\small Mass, radius, central density $\rho_0$ and central pressure $p_0$ for various values of $\xi_0$ for $^4_2$He white dwarfs.}
\centering
\scalebox{0.8}{
\begin{tabular}{c@{\hskip 0.3in}c@{\hskip 0.3in}c@{\hskip 0.3in}c@{\hskip 0.3in}c@{\hskip 0.3in}c}
\specialrule{.1em}{.05em}{.05em} 
\multirow{2}{*}{$\xi_0$} &\multirow{2}{*}{$\displaystyle\left(\frac{p_F}{m_e c}\right)_0$} & Mass &  Radius & $\rho_0$ & $p_0$\\

& &($M_\odot$)& (km) & ($2\times10^{10}$ \ g/cm$^3$) & ($2\times10^{28}$\ dyne/cm$^2$)\\
\specialrule{.1em}{.05em}{.05em} 
$2.35$    & $0.6219$   & $0.2957$ & $12538.45$	 &  $2.3425\times10^{-5}$ & $3.9601\times10^{-7}$\\
$6.20$   & $2.2496$   & $0.9576$ &  $6015.35$   &  $1.1091\times10^{-3}$   & $1.3233\times10^{-4}$ \\
$15.28$  & $22.7911$ & $1.4166$ &  $1029.87$   &  $1.1581$		         & $1.6164$                       \\
$19.00$  & $57.7878$ & $1.3890$ &   $436.95$   &  $19.0131$ 		         & $66.9168$	 	        \\
$\infty$ & $\infty$ & 0.4583  &   52.23                    & $\infty$                             &$ \infty$                         \\
\specialrule{.1em}{.05em}{.05em} 
\end{tabular}
}
\end{table}
To compare these results of numerical integration with the results following from Newtonian gravity, we also numerically integrate the following (Newtonian) equations.
\begin{equation}
\frac{du}{dx}=x^2 \sinh^3\frac{\xi}{4}
\label{eq:eighteen}
\end{equation}
\begin{equation}
\frac{d\xi}{dx}=-\frac{u(x)\sinh^3\frac{\xi}{4}}{(\cosh\xi-4\cosh
\frac{\xi}{2}+3) x^2}
\label{eq:nineteen}
\end{equation}
\begin{figure}[h]
\vskip1cm
\begin{center}
\captionsetup{justification=centering}
\includegraphics[width=3.5in]{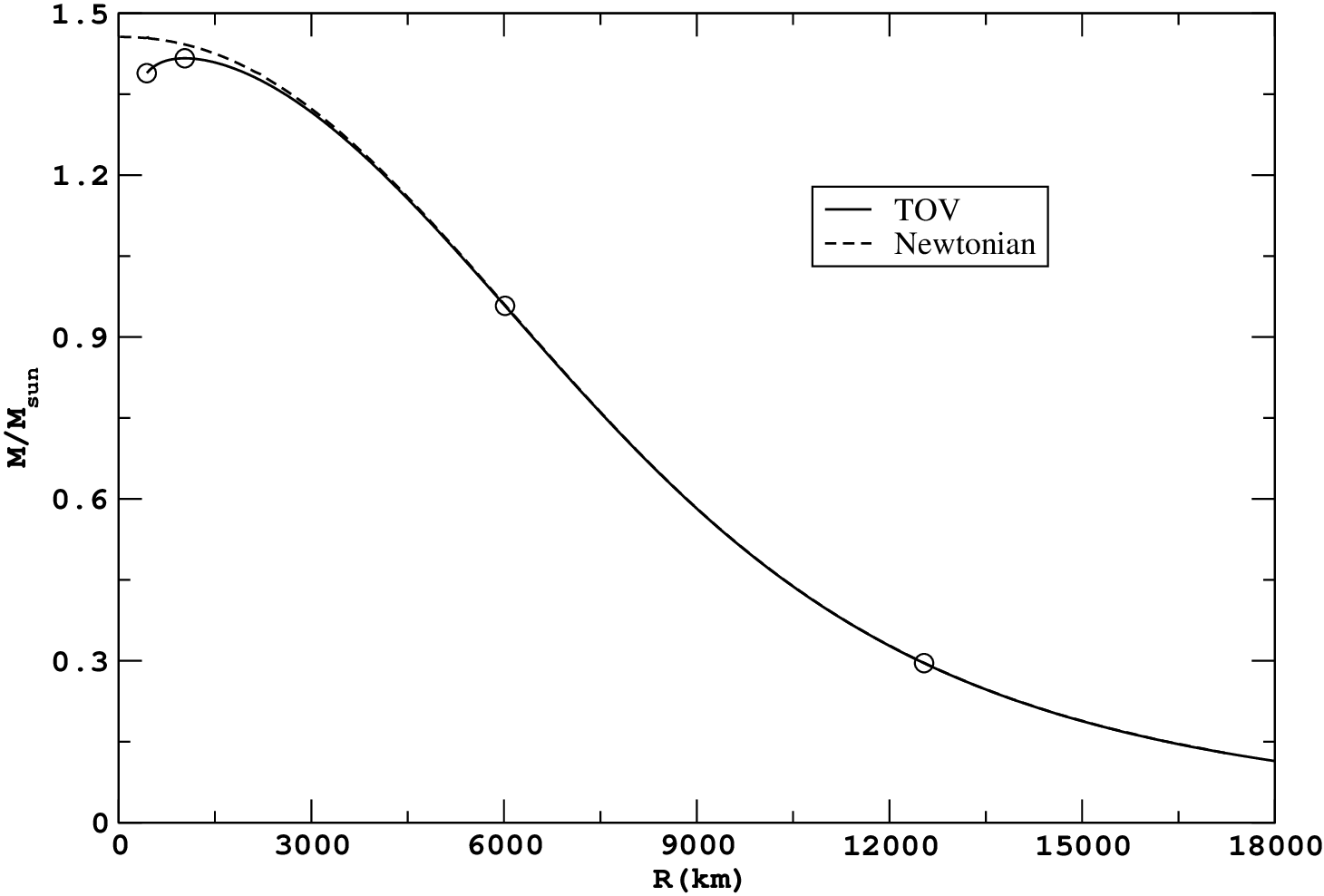}
\caption{\small Mass-Radius relationships given by the TOV (solid curve) and the Newtonian (dashed curve) cases for $^4_2$He white dwarf stars with ideal degenerate EoS. The data-points shown encircled correspond to the first four entries in Table~1.}
\end{center}
\label{a}
\end{figure}
\vskip-0.5cm
The mass-radius relationship for a $^4_2$He white dwarfs following from the numerical integrations of the TOV equations [Eqs.\ (\ref{eq:twelve}) and (\ref{eq:thirteen})] and the Newtonian equations [Eqs.\ (\ref{eq:eighteen}) and (\ref{eq:nineteen})] are compared in Fig.~1. It is seen that the two mass-radius curves coincide for small values of $\xi_0$. This is due to the fact that the TOV equation reduces to the Newtonian equation for small values of central densities as a result of negligible contribution from the internal energy and the smallness of the metric correction $(2GM/c^2 R)$. For small values of $\xi_0$, the equation of state reduces to the form $p=C\rho^{5/3}$ and the mass-radius relation behaves as
$M\sim R^{-3}$ in the the right-hand part of the plot.

For higher values of  central densities, the TOV curve starts to deviate from the Newtonian curve, as seen towards the left-hand part of the plot. Thus, for large
$\xi_0$, there is a departure from the non-relativistic $M\sim R^{-3}$ behaviour for both Newtonian and TOV cases.  We also find that the critical mass is lower for the TOV case than the Newtonian case, the values being $1.4166M_\odot$ and $1.4562M_\odot$ respectively. 
\subsection{Non-ideal case}

A realistic treatment of the electron gas must include various types of interactions among the particles. We therefore consider the Salpeter EoS given by Eqs. (\ref{eq:esalp}) and (\ref{eq:psalp}) and couple them with the TOV equations (\ref{eq:five}) and  (\ref{eq:six}). A comparison between the ideal and the non-ideal cases for the mass-radius relations of $^4_2$He white dwarf stars is shown in Fig.~2. As seen from the plot, the two curves do not coincide for any value of $\xi_0$, although the deviation becomes smaller for higher values of $\xi_0$. As a result, the critical value decreases to $1.4081M_\odot$ which is about 0.6\% lower than the ideal value of $1.4166M_\odot$. We also observe that $^{12}_{6}$C white dwarf star acquires a critical mass of $1.3916M_\odot$ which is different from that of $^4_2$He ($1.4081M_\odot$) when the Salpeter EoS is taken into account. 
Table~2 displays a comparison of  the critical values of masses and radii for $^4_2$He, $^{12}_{6}$C  and $^{56}_{26}$Fe white dwarf stars.
\begin{figure}[h]
\vskip1cm
\begin{center}
\captionsetup{justification=centering}
\includegraphics[width=3.5in]{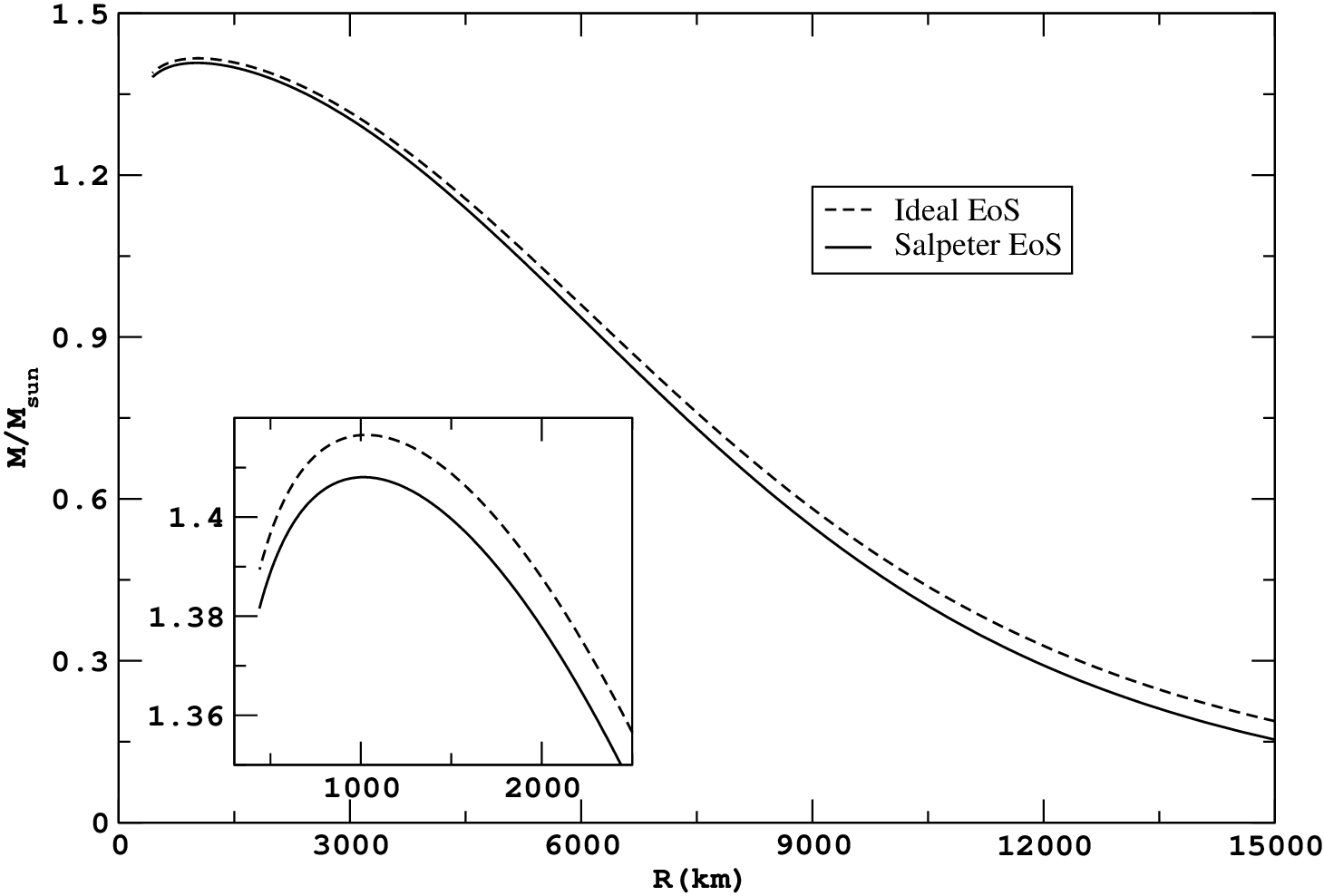}
\caption{\small Mass-Radius relationships given by the TOV equations for $^4_2$He white dwarf stars. The dashed curve represents the solutions with ideal degenerate EoS and the solid curve represents the solutions with Salpeter EoS. The inset shows a magnified view around the region of the maxima in the two cases.}
\label{b}
\end{center}
\end{figure}

We also solve the TOV  equations [Eqs. (\ref{eq:five}) and  (\ref{eq:six})] coupled with the Salpeter EoS  [Eqs. (\ref{eq:esalp}) and (\ref{eq:psalp})] for  $^{16}_{\ 8}$O, $^{20}_{10}$Ne, $^{24}_{12}$Mg,  $^{28}_{14}$Si, $^{32}_{16}$S white dwarf stars. The corresponding results are displayed in the fifth column of Table 3.
\begin{table}[h]
\caption{\small Comparison of  the critical values of masses and radii for $^4_2$He, $^{12}_{6}$C  and $^{56}_{26}$Fe white dwarf stars obtained by solving the TOV equations (\ref{eq:five}) and  (\ref{eq:six}) coupled with the ideal EoS [Eqs. (\ref{eq:eight})] and the Salpeter EoS [Eqs. (\ref{eq:esalp}) and (\ref{eq:psalp})].}
\centering
\scalebox{0.8}{
\begin{tabular}{ccccccc}
\toprule 
   & \multicolumn{3}{c}{Critical mass $(M_\odot)$} & \multicolumn{3}{c}{Critical radius (km)}\\   
    \specialrule{.1em}{.05em}{.05em}
                                    & Ideal EoS      & Salpeter EoS  & \% decrease  &  Ideal EoS & Salpeter EoS  &  \% decrease \\
\specialrule{.1em}{.05em}{.05em}
$^4_2$He             & $1.4166$       &$1.4081$          &  $0.60$     &  $1029.87$ & $1012.07$     &  $1.73$  \\  
$^{12}_6$C          &  $1.4166$      & $1.3916$         &   $1.76$      & $1029.87$   &$1004.59$      & $2.45$   \\     
 $^{56}_{26}$Fe   & $1.2230$       & $1.1565$          & $5.44$      & $927.13$      &	$886.55$     & $4.38$   \\    
\specialrule{.1em}{.05em}{.05em}
\end{tabular}
}
\end{table}

\section{Stability of White Dwarf Stars}

In this section we shall consider the gravitational instability as well as the inverse $\beta$-decay instability for white dwarf stars.

\subsection{Gravitational Instability}

The equilibrium configuration of the star is identified by the extremum in the energy-matter distribution curve. The total energy of the star was previously derived by \cite{shapiro1983} by adding correction due to general relativity, given by $E_T=E_{\rm int}+\Delta E_{\rm int}+E_{\rm Newt}+\Delta E_{\rm GR}$, where the first two terms correspond to the internal energy and the corresponding relativistic correction, the third term is the  gravitational energy in the Newtonian limit, and the fourth term is the correction due to general relativity. The hydrostatic Eqs.\ (\ref{eq:eighteen}) and (\ref{eq:nineteen}), based on Newtonian gravity, ignore the internal kinetic energy contribution and metic correction. Consequently the general relativistic effects are taken into account by adding first order corrections in the above equation as shown. Thus this expression for $E_T$ is an approximate expression for energy due to the general relativistic correction. This approximation is expected to be good when the mass-radius ratio in metric correction is small.

Minimizing this energy $E_T$ gives the equilibrium condition $\frac{dE_T}{d\rho_0}|_{\rho_0=\rho_c}=0$. Minima and maxima correspond to stable and unstable equilibria, given by
$\frac{\partial M}{\partial \rho_0}>0$ and $\frac{\partial M}{\partial \rho_0}<0$, respectively. For $\frac{\partial M}{\partial \rho_0}<0$, the electron degenerate pressure is smaller than the inward gravitational pull causing the star to collapse continuously. The onset of this collapse was obtained by \cite{shapiro1983} by setting $\frac{\partial^2 E_T}{\partial \rho_0^2}=0$. Thus the expression for density for the onset of gravitational instability was calculated as
\begin{equation}
\rho^{\rm ST}_c=2.646\times 10^{10}\left(\frac{\mu_e}{2}\right)^2 \rm g/cm^3 
\label{eq:twentytwo}
\end{equation}
At this critical value of the central density, the star becomes unstable against the gravitational pull. As noted above, this expression for the onset density of gravitational instability is based on the above approximate expression for $E_T$. The critical values  for the central density $\rho^{\rm ST}_c$ for a few white dwarf stars following from Eq.\ (\ref{eq:twentytwo}) are shown in Column~2 of Table~3.

In our present calculations, we take an alternative route to obtain the critical values for the central density, denoted by $\rho_c^{\rm TOVS}$. Thus we obtain the critical central density $\rho_{c}^{\rm TOVS}$ from the solution of the full TOV equations [namely, Eqs.\ (\ref{eq:five}) and (\ref{eq:six}) coupled with Salpter EoS given by Eqs. (\ref{eq:esalp}) and (\ref{eq:psalp})] without making any approximations. Thus, our results are expected to be close to exact.
\begin{figure}
\vskip1cm
\begin{center}
\captionsetup{justification=centering}
\includegraphics[width=3.5in]{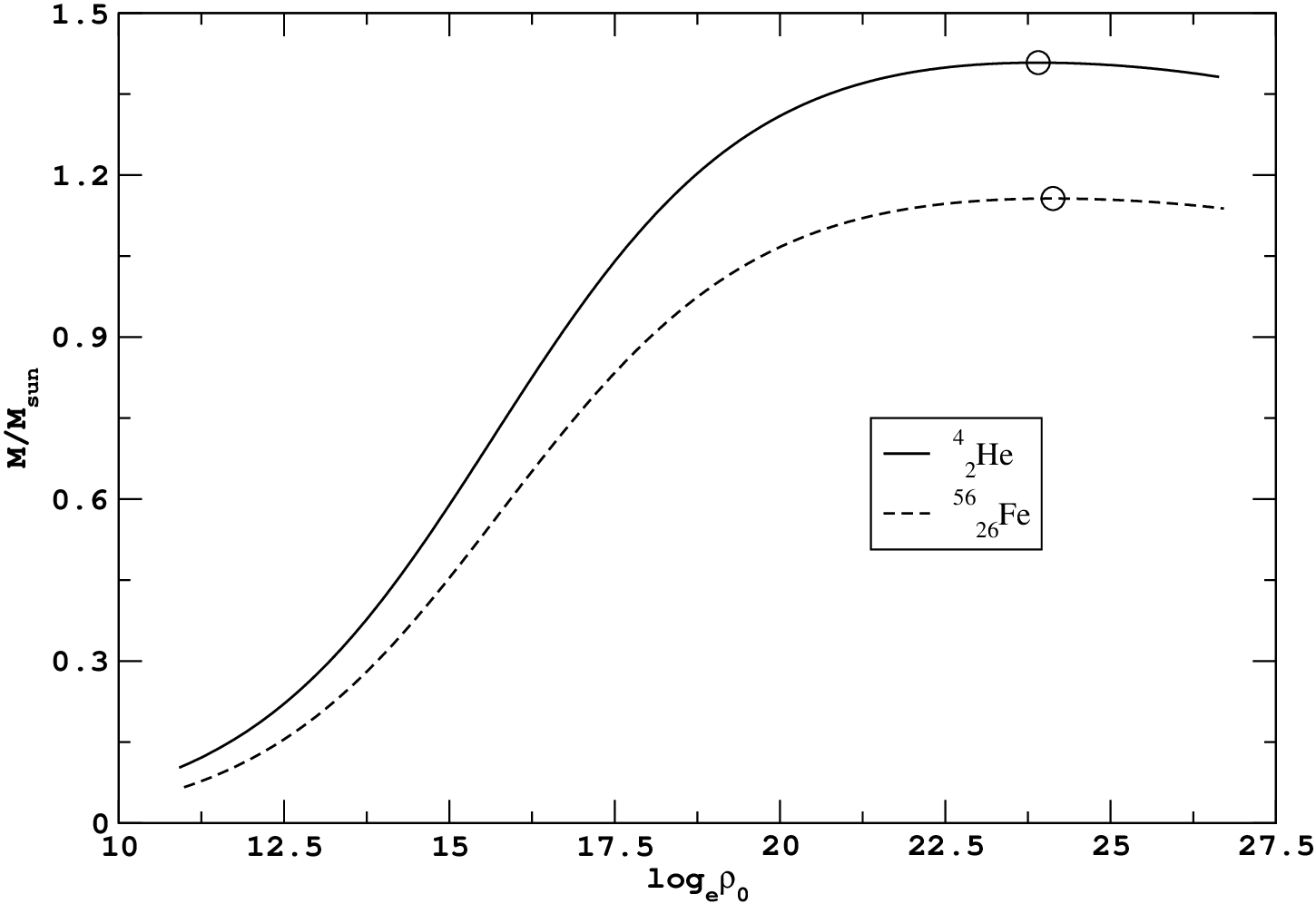}
\caption{\small Plots for $M$ vs $\log_e \rho_0$ for $^4_2$He (solid curve) and  $^{56}_{26}$Fe (dashed curve) white dwarfs obtained by solving the TOV equations (\ref{eq:five}) and (\ref{eq:six}) coupled with Salpeter EoS (\ref{eq:esalp}) and (\ref{eq:psalp}).}
\end{center}
\label{c}
\end{figure}
The plot in Fig.~3 shows the dependence of masses on the central densities $\rho_0$ of $^4_2$He  and $^{56}_{26}$Fe white dwarf stars  as a result of computation based on the TOV equation coupled with Salpter EoS. We see that the mass of the star increases with increase in the central density until a maximum is reached beyond which it falls down. The positive slope ($\frac{\partial M}{\partial \rho_0}>0$) corresponds to the stable portion whereas the negative slope ($\frac{\partial M}{\partial \rho_0}<0$) to the unstable portion. The plot also predicts a maximum stable mass a white dwarf can achieve, which are very nearly $1.4081M_\odot$ for $^4_2$He and $1.1565M_\odot$ for $^{56}_{26}$Fe white dwarfs.  These correspond to central densities of $2.4230\times10^{10}$ g/cm$^3$ for $^4_2$He and $2.9637\times10^{10}$ g/cm$^3$ for $^{56}_{26}$Fe white dwarfs.  For higher values of the densities, the stars become unstable and collapse under their own gravitational pull.

The occurrence of a critical mass can be clearly seen when we plot the mass $M$ versus the radius $R$ of the star on linear scales. This is shown in Fig.~4 for $^4_2$He, $^{12}_6$C and  $^{56}_{26}$Fe white dwarf stars, where the maxima are identified as the critical points, at masses of about $1.4081M_\odot$ for $^4_2$He, $1.3916M_\odot$ for $^{12}_6$C and for $1.1565M_\odot$ $^{56}_{26}$Fe white dwarfs. The portions towards the right of the maxima correspond to stable equilibria, whereas those towards the left correspond to instability.
  
The critical central densities $\rho_c^{\rm TOVS}$ for a few other white dwarf stars ($^{16}_{\ 8}$O, $^{20}_{10}$Ne,  $^{24}_{12}$Mg,  $^{28}_{14}$Si, and $^{32}_{16}$S) are also computed numerically based on the TOV equation [(\ref{eq:five}) and (\ref{eq:six})]  coupled with the Salpeter EoS [(\ref{eq:esalp}) and (\ref{eq:psalp})] and the results are shown in Column~3 of Table~3. In comparison to $\rho_c^{\rm ST}$ we see that the values for $\rho_c^{\rm TOVS}$ are lower in magnitude. The corresponding critical values for the masses $M_c^{\rm TOVS}$  following from the TOV equation coupled with the Salpeter EoS are displayed in Column~4 of Table~3.
\begin{figure}[h]
\begin{center}
\vskip1cm
\captionsetup{justification=centering}
\includegraphics[width=3.5in]{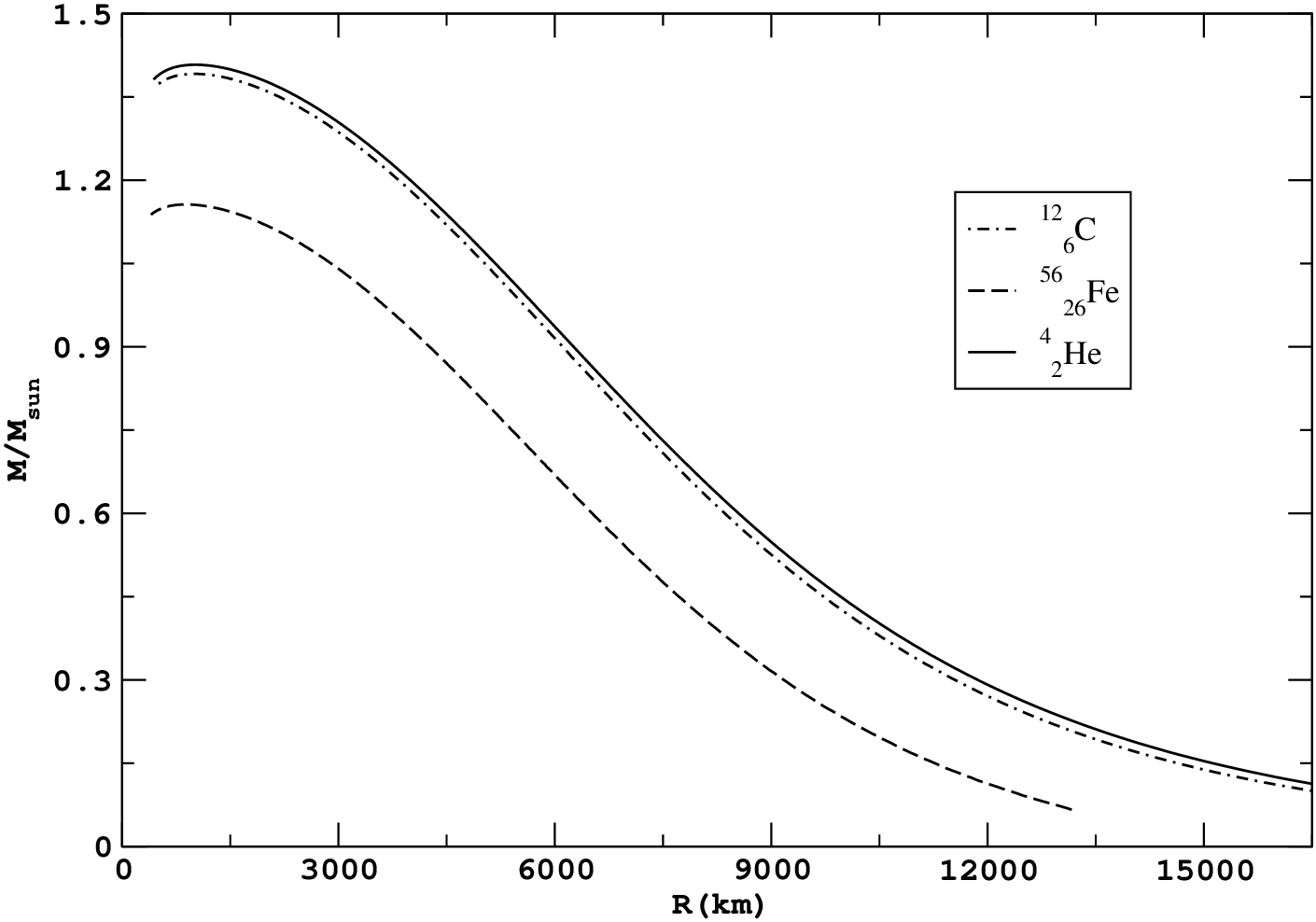}
\caption{\small Mass-radius relations for $^4_2$He (solid curve), $^{12}_6$C (dot-dashed curve) and $^{56}_{26}$Fe (dashed curve) white dwarfs obtained by solving TOV equations (\ref{eq:five}) and (\ref{eq:six}) coupled with Salpeter EoS (\ref{eq:esalp}) and (\ref{eq:psalp}).} 
\label{d}
\end{center}
\vskip-1cm
\end{figure}

\subsection{Inverse $\mathbf{\beta}$-Decay Instability}

The process of inverse $\beta$-decay, namely $^A_ZX + e\longrightarrow ^{\ \ A}_{Z-1}\!\!Y +\nu_e$, becomes important for high values of electron densities. This fact was
ignored when calculating the most stable configurations admitted by the gravity alone. At high electron densities, the electrons become more relativistic so that the condition
$E_F\geqslant \varepsilon_Z$, where $E_F$ is the Fermi energy and $\varepsilon_Z$ is the difference in binding energies of the parent and daughter nuclei, may be
satisfied for inverse $\beta$-decay to occur.

At a sufficiently high density, the star becomes unstable under inverse $\beta$-decay and collapses to form much dense matter (this might be a mixture of neutron rich
nuclei, electrons and neutrons). The threshold density $\rho_{\beta}$ for inverse $\beta$-decay was calculated by \cite{salpeter} by setting $E_F=\varepsilon_Z$. \cite{Rueda} expresses it as
\begin{equation}
\rho_{\beta}=\frac{8\pi\mu_e H}{3h^3 c^3}(\varepsilon_Z^2 + 2m_e c^2\varepsilon_Z)^{3/2}.
\label{eq:twentythree}
\end{equation} 
The $\beta$-decay energy $\varepsilon_Z$ was obtained experimentally by\ \cite{cameron}, as displayed in  Column~4 of Table~3.  The corresponding values of the threshold densities $\rho_{\beta}$ following from Eq.\ (\ref{eq:twentythree}) are shown in Column~5 of Table~3.
\begin{table}[ht!]
\caption{\small Critical values in general relativity and neutronization thresholds for different white dwarfs. Here $\rho^{\rm ST}_c$ is the previous estimate [\cite{shapiro1983}] of the critical central density, $\rho_c^{\rm TOVS}$ is the critical central density obtained from the TOV equations coupled with Salpeter EoS and $\rho_{\beta}$ is the neutronization threshold density. The gravitational critical mass $M_c^{\rm TOVS}$ and the neutronization threshold mass $M_{\beta}^{\rm TOVS}$ follow from the solutions of the TOV equations coupled with Salpeter EoS for the corresponding central densities $\rho_c^{\rm TOVS}$ and $\rho_{\beta}$, respectively.}
\centering
\scalebox{0.8}{
\begin{tabular}{cccccccc}
\specialrule{.1em}{.05em}{.05em} 
& $\rho^{\rm ST}_c$& $\rho_c^{\rm TOVS}$ &$\varepsilon_Z$  & $\rho_{\beta}$& $M_c^{\rm TOVS}$ & $M_{\beta}^{\rm TOVS}$ & \% decrease\\
&($2\times10^{10}$ g/cm$^3$)& ($2\times10^{10}$ g/cm$^3$) & (MeV) & ($2\times10^{10}$ g/cm$^3$) & $(M_{\odot}$)& ($M_{\odot}$)& \\
\specialrule{.1em}{.05em}{.05em} 
$^4_2$He      &$1.3250$& $1.21150$  &  $20.596$ &  $6.85751$ & $1.4081$ &  --- &---\\
$^{12}_{\ 6}$C&$1.3250$& $1.22060$  &  $13.370$ &  $1.94826$ & 1.3916&  --- &---\\
$^{16}_{\ 8}$O&$1.3250$& $1.22979$  &  $10.419$ &  $0.94996$ & 1.3849& $1.3846$&$0.02$\\
$^{20}_{10}$Ne&$1.3250$&$1.26732$  &  $7.026$ &  $0.31036$    & $1.3788$ & $1.3702$&$0.62$\\
$^{24}_{12}$Mg&$1.3250$& $1.26730$  &  $5.513$ &  $0.15784$    & $1.3731$ & $1.3523$&$1.51$\\
$^{28}_{14}$Si&$1.3250$& $1.26728$  &  $4.643$ &  $0.09861$    & $1.3677$ & $1.3341$&$2.46$\\
$^{32}_{16}$S &$1.3250$& $1.26727$  &  $1.710$ &  $0.00370$    & $1.3625$ & $1.1649$&$14.50$\\
$^{56}_{26}$Fe&$1.5289$& $1.48186$  &  $3.695$ &  $0.05720$    & $1.1565$ & $1.0667$&$$7.76\\
\specialrule{.1em}{.05em}{.05em} 
\end{tabular}
}
\end{table}

For white dwarfs whose neutronization density $\rho_{\beta}$ is smaller than the onset density of gravitational instability $\rho_{\rm TOVS}$, we expect an unstable phase before reaching the critical mass $M_c^{\rm TOVS}$ obtained from general relativity. Inverse $\beta$-decay that sets in before the gravitational instability reduces the electron density which in turn reduces the degeneracy pressure so that we expect a smaller value for  the critical mass than that obtained by general relativity. Table~3 compares the critical central densities $\rho_c^{\rm TOVS}$ and inverse $\beta$-decay threshold densities $\rho_\beta$ for white dwarfs of different compositions. 

Comparing the threshold densities given in the third and fifth columns of Table~3, we see, for $^4_2$He and $^{12}_{\ 6}$C white dwarf stars, that gravitational instability sets in before neutronization instability can set in. This implies that the critical mass for $^4_2$He and $^{12}_{\ 6}$C white dwarf stars are $1.4081M_\odot$ and $1.3916M_\odot$, respectively.  For all the rest of the stars, threshold density $\rho_\beta$ for neutronization starts before the critical density $\rho_c^{\rm TOVS}$ for gravitational instability is reached. This implies that the critical masses of $^{16}_{\ 8}$O, $^{20}_{10}$Ne,  $^{24}_{12}$Mg,  $^{28}_{14}$Si, and $^{32}_{16}$S  and $^{56}_{26}$Fe white dwarf stars must be lower than their corresponding critical mass for gravitational instability. We compute the corresponding masses for neutronization thresholds numerically from the TOV equations (\ref{eq:five}) and (\ref{eq:six}) coupled with Salpeter EoS (\ref{eq:esalp}) and (\ref{eq:psalp})  and identify them as the critical masses $M_{\beta}^{\rm TOVS}$ for  stability against neutronization which are shown in the seventh column of Table~3.
\vskip-0.5cm
\section{Conclusions}

 It is seen from our above investigations that the general relativistic effects have lowered the limiting mass of a $^4_2$He white dwarf star. In the ideal degenerate gas approximation, the new value ($1.4166 M_\odot$) obtained by solving the TOV equation is not too far from the Newtonian limit ($1.4562M_\odot$). Thus the general relativistic effects are small in the case of a white dwarf star that lowers the limiting value by approximately 2.7\%.  Our calculated value of $1.4166 M_\odot$ is slightly different from the results obtained by \cite{Bera2016}. They obtained $1.4158M_{\odot}$ and $1.4155M_{\odot}$ in two different general relativistic computations which are approximately 2.36\% and 2.38\% lower than their Newtonian computation value of $1.45M_{\odot}$. 

The equation of state we assumed is valid for all electron velocities, both non-relativistic and ultra relativistic, connecting the two regimes smoothly. Although initially the inter-particle interaction was neglected by assuming an EoS for an ideal degenerate electron gas, we later incorporated it via the Salapeter EoS that includes Coulomb effects, Thomas-Fermi correction, exchange energy and correlation energy. When the TOV equations are solved incorporating these corrections, the critical mass turned out to be $1.4081$ for $^4_2$He and  $1.3916$ for $^{12}_6$C white dwarfs which are slightly lower than the ideal value ($1.4166M_{\odot}$). We have shown these differences in Table 2 that also includes the values for  $^{56}_{26}$Fe white dwarf. The maxima of the mass-radius curve (in Fig. \ref{d} and Fig. \ref{b}) mark the onset of gravitational collapse and the regions towards the left of the maxima correspond to unstable regions. It is expected that the growth in density makes the electrons more relativistic so that the condition favouring inverse beta decay is approached.

We took account of  neutronization by using Eq.\ (\ref{eq:twentythree}) obtained on the basis of Salpeter's arguments. The onset density $\rho_\beta$ so obtained for inverse $\beta$-decay for different compositions of the star are shown in Column~5 of Table~3. A comparison with $\rho_c^{\rm TOVS}$ (shown in Column~3 of Table~3) indicates that stars composed of lighter elements ($^4_2$He and $^{12}_{\ 6}$C) are more stable than those composed of heavier elements ($^{16}_{\ 8}$O, $^{20}_{10}$Ne,  $^{24}_{12}$Mg,  $^{28}_{14}$Si, $^{32}_{16}$S and $^{56}_{26}$Fe). The onset of inverse $\beta$ decay is found to start before reaching the onset of gravitational instability in the white dwarfs composed of heavier elements. The maximum stable mass $M_c^{\rm TOVS}$ obtained by solving the TOV equations coupled with Salpeter EoS for these stars (as given in column~6 of Table~3) no longer can be identified as their critical masses. Consequently for these stars, we solve the TOV equations coupled with Salpeter EoS corresponding to the central densities $\rho_\beta$ to obtain the maximum stable mass $M_\beta^{\rm TOVS}$ as shown in the seventh column of Table~3. 

Mass distribution of a large number of white dwarf stars with a wide range of masses, including low and massive stars, were plotted by  \cite{bergeron} and \cite{kepler}. The most massive non-magnetic white dwarf observed was LHS4033 [\cite{dahn, bergeron, kepler}] which was predicted to have an oxygen-neon core with a mass in the range of $1.318$---$1.335M_{\odot}$. This range of mass values is compatible with our calculations as we see from the seventh column of Table 3. 

However, recent observations of type Ia supernova (SNe Ia) admit white dwarfs with masses as high as $2.3-2.6M_{\odot}$. \cite{Howell2006} argued that the over-luminosity and low expansion velocities around SN 2003fg white dwarf could be explained if it assumed to have a mass greater than $1.44M_{\odot}$. \cite{Hicken2007} presented SN2006gz as a possible SNe Ia candidate that was identified with similar properties. \cite{Scalzo2010} estimated the total mass of  SN 2007if progenitor to be in the range $2.2-2.6M_{\odot}$. \cite{Silverman2011} suggested another member of SNe Ia class, SN2009dc,  with similar peculiarities, possibly formed from the merger of two white dwarfs. 

It has been speculated that the presence of a magnetic field drastically modifies the situation in a compact star due to the Landau quantization of the electronic energy levels.
\cite{Gao2011} simulated the electron $\beta$-capture in a magnetar by considering electrons belonging to the higher Landau levels in a super high magnetic field that admit the energy threshold values for inverse $\beta$-decay to occur. For a $^4_2$He white dwarf star, \cite{Das2012} considered the modified equation of state due to the Landau levels of the electrons in a magnetic field. They showed that a maximum mass of $2.3M_{\odot}$ is reached at the highest turning point in the mass-radius relation of a $^4_2$He white dwarf star within the Newtonian (Lane-Emden) framework. They indicated that the presence of high magnetic fields can give rise to white dwarfs of masses as high as $2.3-2.6M_{\odot}$ with radii around $600$ km. It is interesting to mention that \cite{Gao2013} and \cite{Zhu2016} made a great deal of deliberations in this direction, particularly in the study of how the Fermi energy and the electron degeneracy pressure change in a neutron star due to the presence of a strong magnetic field. 

\cite{Gao2013} and \cite{Li2016} indicated that in the presence of a strong magnetic field the electron degeneracy pressure increases with strength of the magnetic field. This is because the pressure is proportional to $E_F^4$ and the Fermi energy $E_F$ increases with the strength of the magnetic field. Consequently, the number density $n_e$ (and hence the matter density), being determined by the Fermi energy, also increases with the magnetic field. Thus, with respect to the non-magnetic case, a greater matter density near and away from the centre could be supported against the gravitational pull by the increased pressure gradient in the presence of a strong magnetic field. 
 
It may also be remarked that soft gamma ray repeaters (SGRs) and anomalous X-ray pulsars (AXPs) were thought to be magnetars until a recent exception was observed in SGR 0418+5729 which was found to be inconsistent with the magnetar model of SGRs and AXPs based on neutron stars. \cite{Malheiro2012} showed that the observed upper limit on the spin down rate of SGR 0418+5726 is in accordance with a model based on a massive fast rotating highly magnetised white dwarf.  It may thus be speculated that their masses would be higher than $1.4562M_{\odot}$ due to the effect of strong magnetic field.

\subsection*{Acknowledgement}
M.K.N. is indebted to Indian Institute of Technology Delhi for their kind hospitality during his frequent visits.

\end{document}